\documentclass[11pt]{article}

\usepackage{amssymb,amsmath,amsthm,epsfig}

\theoremstyle{plain}
\newtheorem{prop}{Proposition}[section]
\newtheorem{thm}[prop]{Theorem}

\newtheorem{lem}[prop]{Lemma}

\theoremstyle{remark}

\newcommand{\ovl}{\overline}
\newcommand{\n}{\noindent}

\newcommand{\wt}{\widetilde}

\begin{document}

\title{Generalization of Grover's Algorithm to Multiobject\\ Search in Quantum 
Computing, Part II: \\ General Unitary Transformations}

\author{Goong Chen$^{(1),(2)}$ and Shunhua Sun$^{(3)}$}

\date{}
\maketitle

\abstract{There are  major advantages in a newer version of Grover's quantum 
algorithm [\ref{G3}] utilizing a general unitary transformation in the search of 
a single object in a large unsorted database. In this paper, we generalize this 
algorithm to multiobject search. We show the techniques to achieve the reduction 
of the problem to one on an invariant subspace of dimension just equal to two.}

\vfil

\begin{itemize}
\item[(1)] Department of Mathematics, Texas A\&M University, College Station, TX 
\ 77843-3368. E-mail: \ gchen@math.tamu.edu.
\item[(2)] Supported in part by Texas A\&M University Interdisciplinary Research 
Initiative IRI 99-22.
\item[(3)] Department of Mathematics, Sichuan University, Chengdu, Sichuan, 
China. Supported in part by a grant from Natural Science Foundation of China.
\end{itemize}

\eject

\baselineskip = 18pt

\section{Introduction}\label{gen:sec1}

\indent

This paper is a continuation from [\ref{C1}] on quantum computing algorithms for 
multiobject search.

L.K.~Grover's first papers [\ref{G1}, \ref{G2}] on ``quantum search for a needle 
in a haystack'' have stimulated broad interest in the theoretical development 
of quantum computing algorithms. Let an unsorted database consist of $N$ objects 
$\{w_j\mid 1\le j\le N\}$; each object $w_j$ is stored in a quantum computer 
(QC) memory as an eigenstate $|w_j\rangle$, $j=1,2,\ldots, N$, with $\mathcal{B} 
\equiv \{|w_j\rangle\mid 1\le j\le N\}$ forming an orthonormal basis of a 
Hilbert space $\mathcal{H}$. Let $|w\rangle$ be an element of $\mathcal{B}$ 
which is the (single) object to be searched. Grover's algorithm in [\ref{G1}, 
\ref{G2}] is to utilize a unitary operator
\begin{equation}\label{gen:eq1.1}
U \equiv -I_sI_w
\end{equation}
where
\begin{align}
\label{gen:eq1.2}
I_w &\equiv \pmb{I} -2|w\rangle \langle w|, \qquad (\pmb{I} \equiv \text{ the 
identity operator on } \mathcal{H})\\
\label{gen:eq1.3}
I_s &\equiv \pmb{I} -2|s\rangle \langle s|,\qquad |s\rangle \equiv \frac1{\sqrt 
N} \sum^N_{i=1} |w_i\rangle,
\end{align}
to perform the iterations $U^m|s\rangle$, which will lead to the target state 
$|w\rangle$ with probability close to 1 after approximately $\frac\pi4 \sqrt N$ 
number of iterations. The algorithm is of optimal order.

In a more recent paper [\ref{G3}], Grover showed that the state $|s\rangle$ in 
(\ref{gen:eq1.3}) can be replaced by {\em any\/} quantum state 
$|\gamma\rangle$ with nonvanishing amplitude for each object $w_j$ and, 
correspondingly, the Walsh-Hadamard operator previously used by him to construct 
the operator $I_s$ can be replaced by a sufficiently general nontrivial unitary 
operator. Grover's new ``search engine'' in [\ref{G3}] is a unitary operator 
taking the form
\begin{equation}\label{gen:eq1.4}
U = -I_\gamma V^{-1}I_wV\colon \ \mathcal{H}\to \mathcal{H}
\end{equation}
where $V$ is an {\em arbitrary\/} unitary operator. The object $w$ will be 
attained (with probability close to 1) by iterating $U^m|\gamma\rangle$.

This seems to give the algorithm/software designer large flexibility in 
conducting quantum computer search and code development. It increases the 
variety of quantum computational operations that can feasibly be performed by 
practical software. In particular, it opens the possibility of working with an 
initial state $|\gamma\rangle$ (in place of $|s\rangle$) that is other than a 
superposition of exactly $N=2^n$ ($n=$ number of qubits) alternatives. This 
suggests a new paradigm in which the whole dataset (not just the key) is encoded 
in the quantum apparatus. This new point of view may also overcome some of the 
practical difficulties noted by Zalka [\ref{Z1}] in searching a physical 
database by Grover's method.

In the next section, we study the generalization of (\ref{gen:eq1.4}) to 
multiobject search. 

\section{Multiobject Search Algorithm Using a General Unitary 
Transformation}\label{gen:sec2}

\setcounter{equation}{0}

\indent

Let $\{|w_i\rangle\mid 1\le i\le N\}$ be the basis of orthonormal eigenstates 
representing an unsorted database $w_i$, $1\le i\le N$, as noted in \S I. We 
inherit much of the notation in [\ref{C1}]: \ let $f$ be an oracle function such 
that
$$f(w_i) = \left\{\begin{array}{ll}
1,&1\le i \le \ell,\\ 0,&\ell+1\le i\le N,\end{array}\right.$$
where $w_i$, $i=1,2,\ldots, \ell$, represent the multiobjects under search. We 
wish to find at least one $w_i$, for $i=1,2,\ldots,\ell$. Let $|\gamma\rangle$ 
be {\em any\/} unit vector in $\mathcal{H}$, and let $L\equiv 
\text{span}\{|w_i\rangle \mid 1\le i \le \ell\}$. 
Define
$$I_\gamma = \pmb{I}-2|\gamma\rangle \langle \gamma|\colon \ \mathcal{H}\to 
\mathcal{H},$$
and
$$I_L|w_j\rangle = (-1)^{f(w_j)} |w_j\rangle, \qquad j=1,2,\ldots, N,$$
and $I_L$ is then uniquely extended linearly to all $\mathcal{H}$ with the 
representation
$$I_L = \pmb{I}-2\sum^\ell_{i=1} |w_i\rangle \langle w_i|.$$
Both $I_\gamma$ and $I_L$ are unitary operators. Let $V$ be any unitary operator 
on $\mathcal{H}$. Now, define
\begin{equation}\label{gen:eq2.1}
U = -I_\gamma V^{-1} I_{L} V.
\end{equation}
Then $U$ is a unitary operator; it degenerates into Grover's operator $U$ in
(\ref{gen:eq1.4}) when $\ell=1$ and further into the old Grover's operator $U$ 
in (\ref{gen:eq1.1}) if $V\equiv \pmb{I}$.

The unit vector $|\gamma\rangle \in \mathcal{H}$ is arbitrary  except that 
we require $V|\gamma\rangle \notin L$. (Obviously, any $|\gamma\rangle$ such 
that $\langle w_i|\gamma\rangle \ne 0$ for all $i=1,2,\ldots, N$, will work, 
including $|\gamma\rangle\equiv |s\rangle$ in (\ref{gen:eq1.3}).) If 
$V|\gamma\rangle \in L$, then
$$V|\gamma\rangle = \sum^\ell_{j=1} g_i|w_i\rangle,\quad g_i\in \mathbb{C}, 
\quad \sum^\ell_{j=1} |g_i|^2 =1.$$
A measurement of the state $V|\gamma\rangle$ will yield an eigenstate 
$|w_j\rangle$, for some $j\colon \ 1\le j \le\ell$, with probability $|g_j|^2$. 
Thus the search task would have been completed. Thus, let us consider the 
nontrivial case $V|\gamma\rangle \notin L$. This implies $|\gamma\rangle \notin 
V^{-1}(L)$ and, hence,
\begin{equation}\label{gen:eq2.2}
\wt L \equiv\hbox{span}(\{|\gamma\rangle\} \cup V^{-1}(L))\
\end{equation}
is an $(\ell+1)$-dimensional subspace of $\mathcal{H}$. It effects a reduction 
to a lower dimensional invariant subspace for the operator $U$, according to the 
following.

\begin{lem}\label{gen:lem2.1}
Assume that $\langle \gamma|\gamma\rangle =1$ and $V|\gamma\rangle\notin L$. 
Then $U(\wt L) = \wt L$. 
\end{lem}

\begin{proof}
For any $j\colon \ 1\le j \le \ell$, denote
$$\mu_{\gamma,j} = \langle w_j|V|\gamma \rangle.$$
\begin{itemize}
\item[(1)] We have, for $j\colon \ 1\le j \le \ell$,
\begin{align}
U(V^{-1}|w_j\rangle) &= -I_\gamma V^{-1}\left(I - 2\sum^\ell_{i=1} |w_i\rangle 
\langle w_i|\right) |w_j\rangle\nonumber\\
&= -I_\gamma V^{-1}(-|w_j\rangle)\nonumber\\
&= I_\gamma V^{-1} |w_j\rangle\nonumber\\
&= (I-2|\gamma\rangle \langle \gamma|) V^{-1}|w_j\rangle\nonumber\\
&= V^{-1}|w_j\rangle -2(\langle \gamma|V^{-1}|w_j\rangle) 
|\gamma\rangle\nonumber\\
\label{gen:eq2.3}
&= V^{-1}|w_j\rangle -2 \ovl\mu_{\gamma,j} \gamma \in \wt L;
\end{align}
\item[(2)]
\begin{align}
U|\gamma\rangle &= -I_\gamma V^{-1} \left(I - 2 \sum^\ell_{i=1} |w_i\rangle 
\langle w_i|\right) (V|\gamma\rangle)\nonumber\\
&= -(I-2|\gamma\rangle \langle\gamma|) \left[|\gamma\rangle - 2\sum^\ell_{i=1} 
(\langle w_i|V|\gamma\rangle) V^{-1}|w_i\rangle\right]\nonumber\\
&= |\gamma \rangle + 2 \sum^\ell_{i=1} \mu_{\gamma,i} V^{-1} |w_i\rangle - 4 
\sum^\ell_{i=1} \mu_{\gamma,i} \ovl\mu_{\gamma,i} |\gamma\rangle\nonumber\\
\label{gen:eq2.4}
&= \left(1- 4 \sum^\ell_{i=1} |\mu_{\gamma,i}|^2\right) |\gamma\rangle + 
2\sum^\ell_{i=1} \mu_{\gamma,i} V^{-1}|w_i\rangle \in \wt L.\qed
\end{align}
\end{itemize}
\renewcommand{\qed}{}\end{proof}
By Lemma~\ref{gen:lem2.1}, the Hilbert space $\mathcal{H}$ admits an orthogonal 
direct sum decomposition
$$\mathcal{H} = \wt L \oplus \wt L^\bot$$
such that $\wt L^\bot$ is also an invariant subspace of $U$. In our subsequent 
iterations, the actions of $U$ will be restricted to $\wt L$, as the following 
Lemma~\ref{gen:lem2.2} has shown. Therefore we can ignore the complementary 
summand space $\wt L^\bot$.

\begin{lem}\label{gen:lem2.2}
Under the same assumptions as Lemma \ref{gen:lem2.1}, we have $U^m|\gamma\rangle 
\in \wt L$ for $m\in \mathbb{Z}^+ \equiv \{0,1,2,\ldots\}$.
\end{lem}

\begin{proof}
It follows obviously from by (\ref{gen:eq2.2}) and Lemma \ref{gen:lem2.1}.
\end{proof}

Consider the action of $U$ on $\wt L$. Even though $|\gamma\rangle, 
V^{-1}|w_i\rangle$, $i=1,\ldots,\ell$, form a basis of $\wt L$, these vectors 
are not mutually orthogonal. We have
\begin{align}\label{gen:eq2.5}
U \left[\begin{matrix}
|\gamma\rangle\\ V^{-1}|w_1\rangle\\ V^{-1}|w_2\rangle\\ \vdots\\ 
V^{-1}|v_\ell\rangle\end{matrix}\right] &= \left[\begin{matrix}
1 - 4 \sum\limits^\ell_{i=1} |\mu_{\gamma,j}|^2&2\mu_{\gamma,1}&2\mu_{\gamma,2}& 
\ldots&2\mu_{\gamma,\ell}\\
-2\ovl \mu_{\gamma,1}&1&0&\ldots&0\\
-2\ovl\mu_{\gamma,2}&0&1&&0\\
\vdots&\vdots&&\ddots&\vdots\\
-2\ovl\mu_{\gamma,\ell}&0&0&\ldots&1\end{matrix}\right] \left[\begin{matrix}
|\gamma\rangle\\ V^{-1}|w_1\rangle\\ V^{-1}|w_2\rangle\\ \vdots\\ 
V^{-1}|w_\ell\rangle\end{matrix}\right],\\
&\equiv \mathcal{M} \left[\begin{matrix}
|\gamma\rangle\\ V^{-1} |w_1\rangle\\ V^{-1}|w_2\rangle\\ \vdots\\
V^{-1}|w_\ell\rangle\end{matrix}\right],\nonumber
\end{align}
according to (\ref{gen:eq2.3}) and (\ref{gen:eq2.4}). Therefore, with respect
to the basis $\{|\gamma\rangle, V^{-1}|w_i\rangle\mid i=1,\ldots,\ell\}$, the
matrix representation of $U$ on $\tilde L$ is $\mathcal{M}^T$, the transpose
of $\mathcal{M}$. These two $(\ell+1) \times (\ell+1)$ matrices $\mathcal{M}$
and $\mathcal{M}^T$ are {\em  nonunitary}, however, because the basis
$\{|\gamma\rangle, V^{-1}|w_i\rangle|$,  $i=1,2,\ldots, \ell\}$ is not
orthogonal. This fact is relatively harmless here, as we can further effect a
reduction of  dimensionality by doing the following. Define a unit vector
\begin{equation}\label{gen:eq2.6} |\mu\rangle = 2\sum^\ell_{j=1}
\mu_{\gamma,j} V^{-1}|w_j\rangle/a,\quad a \equiv  \left(4 \sum^\ell_{j=1}
|\mu_{\gamma,j}|^2\right)^{1/2} >0. \end{equation}

\begin{thm}\label{gen:thm2.3}
Let $\mathcal{V} \equiv \text{span}\{|\gamma\rangle, |\mu\rangle\}$. Then 
$\mathcal{V}$ is a two-dimensional invariant subspace of $U$. We have 
\begin{equation}\label{gen:eq2.7}
U\left[\begin{matrix} |\gamma\rangle\\ |\mu\rangle\end{matrix}\right] = M 
\left[\begin{matrix} |\gamma\rangle\\ |\mu\rangle\end{matrix}\right],\qquad M 
\equiv \left[\begin{matrix} 1-a^2&a\\ -a&1\end{matrix}\right].
\end{equation}
Consequently, with respect to the basis $\{|\gamma\rangle, |\mu\rangle\}$ in
$\mathcal{V}$, the matrix representation of $U$ is $M^T$.
 \end{thm}

\begin{proof}
Using (\ref{gen:eq2.3}), we have
\begin{align*}
U|\mu\rangle &= 2 \sum^\ell_{j=1} \mu_{\gamma,j} V^{-1}|w_j\rangle \cdot \frac1a 
-2 \sum^\ell_{j=1} |\mu_{\gamma,j}|^2\cdot \frac1a |\gamma\rangle\\
&= |\mu\rangle -a|\gamma\rangle.
\end{align*}
Again, from the definition of $|\mu\rangle$ in (\ref{gen:eq2.6}), we see that 
(\ref{gen:eq2.4}) gives
$$U|\gamma\rangle = (1-a^2) |\gamma\rangle + a|\mu\rangle.$$
Therefore (\ref{gen:eq2.7}) follows.
\end{proof}

Theorem \ref{gen:thm2.3} gives a dramatic reduction of dimensionality to {\bf 
2}, i.e., the dimension of the invariant subspace $\mathcal{V}$. Again, we note 
that the matrices $M$ and $M^T$ in (\ref{gen:eq2.7}) are {\em not 
unitary}.

Any vector $|v\rangle \in\mathcal{V}$ can be represented as
$$|v\rangle = c_1|\gamma\rangle +c_2|\mu\rangle,$$
and so
\begin{align*}
U|v\rangle &= U(c_1|\gamma\rangle + c_2|\mu\rangle\\
&= c_1[(1-a^2)|\gamma\rangle + a|\mu\rangle] + c_2[-a|\gamma\rangle + 
|\mu\rangle],
\end{align*}
and thus
\begin{equation}\label{gen:eq2.8}
U|v\rangle = M^T \left[\begin{matrix} c_1\\ c_2\end{matrix}\right] = 
\left[\begin{matrix} 1-a^2&-a\\ a&1\end{matrix}\right] \left[\begin{matrix} 
c_1\\ c_2\end{matrix}\right],
\end{equation}
where the first component of the vector on the right hand side of 
(\ref{gen:eq2.8}) corresponds to the coefficient of $|\gamma\rangle$ while the 
second component corresponds to the coefficient of $|\mu\rangle$. Therefore
\begin{equation}\label{gen:eq2.9}
U^m|\gamma\rangle = \left[\begin{matrix} 1-a^2&-a\\ a&1\end{matrix}\right]^m 
\left[\begin{matrix} 1\\ 0\end{matrix}\right].
\end{equation}
The above can be viewed geometrically ([\ref{J}]) as follows:

\begin{figure}[htpb]
\begin{center}
\epsfig{figure=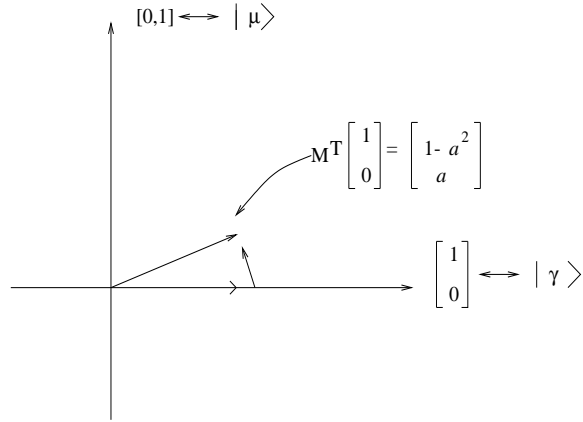,height=2.2in,width=3in}
\end{center}
\caption{A geometric view of a single iteration (\ref{gen:eq2.8})}
\end{figure}

\n $M^T\left[\begin{smallmatrix} 1\\ 0\end{smallmatrix}\right] = 
\left[\begin{smallmatrix} 1-a^2\\ a\end{smallmatrix}\right]$, for $a>0$ very 
small, $a\approx \sin a$, and therefore $\left[\begin{smallmatrix} 1-a^2\\ 
a\end{smallmatrix}\right]$ is a vector obtained from the unit vector 
$\left[\begin{smallmatrix} 1\\ 0\end{smallmatrix}\right]$ by rotating it 
counterclockwise with angle $a$. It takes approximately
$$m \approx \frac{\pi/2}a = \frac\pi{2a} = \pi \bigg/ 4 \left[\sum^\ell_{j=1} 
|\mu_{\gamma,j}|^2\right]^{1/2}$$
rotations to closely align the vector $U^m |\gamma\rangle$ with $|\mu \rangle 
\in V^{-1}L^1$. Thus $V(U^m|\gamma\rangle)$ deviates little from the subspace $L 
= \text{span}\{|w_i\rangle\mid i=1,2,\ldots, \ell\}$. A measurement of 
$VU^m|\gamma\rangle$ gives one of the eigenstates $|w_j\rangle$, for some 
$j\colon \ 1\le j \le\ell$, with probability nearly equal to 1, and the task of 
multiobject search is completed with this large probability.

\end{document}